\documentclass[conference]{IEEEtran}
\usepackage{graphicx,amsmath,amssymb}
\usepackage{subfigure}
\usepackage{citesort}
\usepackage{fancyhdr}
\usepackage{mdwmath}
\usepackage{mdwtab}
\usepackage{balance}
\usepackage{xcolor}
\usepackage{bm}
\usepackage{bbm}
\usepackage{stfloats}
\usepackage{cite}
\usepackage{epstopdf}
\usepackage{algorithm}
\usepackage{algorithmic}
\usepackage{multirow}
\usepackage{flafter}

\usepackage{amssymb}
\usepackage{amsmath}

\usepackage[
top    = 0.7 in,
bottom = 1.05 in,
left   = 0.63 in,
right  = 0.63 in]{geometry}

\newtheorem{theorem}{Theorem}

\newtheorem{lemma}{Lemma}

\newtheorem{corollary}{Corollary}

\begin{document}
\title{Semi-Integrated-Sensing-and-Communication (Semi-ISaC) Networks Assisted by NOMA }

\author{
\IEEEauthorblockN{Chao~Zhang\IEEEauthorrefmark{1}, Wenqiang~Yi\IEEEauthorrefmark{1}, Yuanwei~Liu\IEEEauthorrefmark{1}} \IEEEauthorblockA{\IEEEauthorrefmark{1}Queen Mary University of London, London, UK} }
\maketitle

\begin{abstract}
 This paper investigates non-orthogonal multiple access (NOMA) assisted integrated sensing and communication (ISaC) networks. Compared to the conventional ISaC networks, where the total bandwidth is used for both the radar detection and wireless communications, the proposed Semi-ISaC networks allow that a portion of bandwidth is used for ISaC and the rest of the bandwidth is only utilized for wireless communications. We first derive the analytical expressions of the outage probability for the communication signals, including the signals for the radar target and the communication transmitter. Additionally, we derive the analytical expressions of the ergodic radar estimation information rate (REIR) for the radar echoes. The simulation results show that 1) NOMA ISaC has better spectrum efficiency than the conventional ISaC; and 2) The REIR is enhanced when we enlarge the density of pulses.
\end{abstract}

\section{Introduction}
Thanks to the applications of the fifth generation wireless systems (5G), the available bandwidth of communications has been expanded from the conventional radio frequency (RF) to millimeter waves, which is able to serve a tremendous amount of devices. However, as radar sensing technologies generally consider millimeter waves, the bandwidth of both sensing and wireless communications is gradually overlapped \cite{ISAC_basic2}. It is predicted that in the future communication environments, radar and wireless communication technologies are likely to be more and more similar \cite{ISAC_basic,ISACJ}. Hence, for the sixth generation wireless communications (6G), integrated sensing and communication (ISaC) is an emergency and promising technology as the interference between radars and wireless communication devices is desired to be solved \cite{zhaolin}.

In practical scenarios, it is difficult to utilize all the spectrum for the ISaC use as the bandwidth has been occupied by different applications, including the spectrum of sub-6 GHz band and millimeter waves \cite{band}. For example, there are some applications based on radar and sensing: the L-band (1-2 GHz) for long-range air traffic control and long-range surveillance; the S-band (2-4 GHz) for terminal air traffic control, moderate-range surveillance, and long-range weather resolution; the C-band (4-8 GHz) for long-range tracking, weapon location, and weather observation; and the millimeter waves for high resolution mapping, satellite altimetry, vehicle radars, and police radars \cite{band}. Hence, the most practical scenario is that a portion of the bandwidth is used for ISaC to fit the specific applications while the total bandwidth can be exploited for wireless communications. In a word, compared to the conventional ISaC scenarios, a semi-integrated-sensing-and-communication (Semi-ISaC) case is more practical and promising for future 6G applications.

To achieve Semi-ISaC, compared to orthogonal multiple access (OMA), non-orthogonal multiple access (NOMA) is a good solution as the successive interference cancellation (SIC) technology fits the ISaC scenarios well \cite{xidong}. There are several obvious advantages of utilizing NOMA into the Semi-ISaC networks. Firstly, with the aid of NOMA, the spectrum efficiency is enhanced because each resource block is split to serve more than one user. Additionally, as the study of the SIC scheme is mature, we have a strong basis for fundamental analysis to harness NOMA in ISaC networks \cite{yuanweiNOMA}. Moreover, for the conventional ISaC networks, the base-station (BS) may need predicted radar echoes to enhance the accuracy of computing \cite{ref}, which is to enlarge the difference of power levels between communication signals and radar signals (thus easier to be split by SIC). When we exploit the NOMA technique with power allocation \cite{yuanweiNOMA,powerallocation1}, the transmit powers of the users (for communication signals) and that of the BS (for the radar signals) are able to be well-designed, which releases the accuracy of radar prediction and provides a new degree of freedom for ISaC networks.

Motivated by the advantages above, we investigate a NOMA-assisted Semi-ISaC system. The main contributions are summarized as follows: 1) We present the idea of the NOMA-assisted Semi-ISaC networks; 2) We derive the analytical results of outage probability by closed-form expressions for communication signals for both a radar target and communication transmitter; and 3) We propose and derive a new performance matric called ergodic \emph{Radar Estimation Information Rate (REIR)} as closed-form expressions to investigate the radar echoes.

\begin{figure}
\vspace{-0.2cm}
\centering
\includegraphics[width= 3.1in]{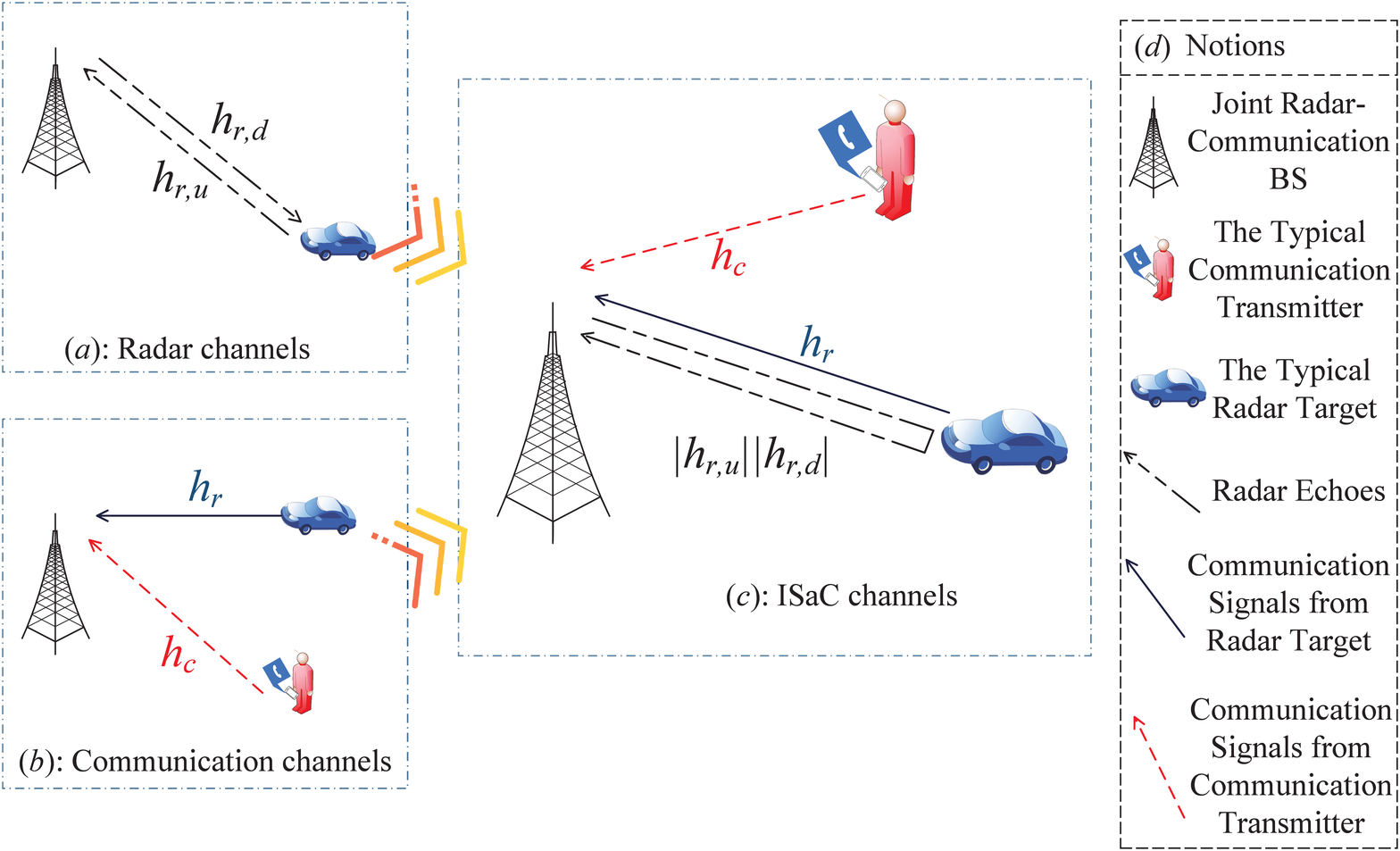}
\caption{The illustration of the channels in ISaC systems: (a) Conventional radar channel; (b) Conventional NOMA channels; (c) NOMA-aided Semi-ISaC channels; and (d) notions.}
\vspace{-0.2cm}
\label{system}
\end{figure}
\section{System model}

We focus on an uplink NOMA-assisted Semi-ISaC system, which includes a BS, a communication transmitter, and a radar target. The total bandwidth is utilized for wireless communication while a portion of the bandwidth is used for the ISaC case with the percentage $\beta_{semi} \in[0,1]$. In this paper, a single-input-single-output (SISO) case is considered. We assume the radar targets also have communication functions, e.g., vehicles (cars or unmanned aerial vehicles). More specifically, the radar echoes reflected from the radar targets and the uplink signals of the communication transmitter are integrated. Thus, the BS receives the radar echoes and two categories of communication signals (from radar targets and communication transmitters) instantaneously. In the following, we present the channel propagation and the NOMA designs for the NOMA-assisted Semi-ISaC system

\subsection{Channel Propagation Gain}

\subsubsection{Small-Scale Fading}
For both radar and communication links, the path loss model and small-scale fading model are defined in this subsection. As the ISaC channels are considered in mmWave bandwidth, we assume that Nakagami-\emph{m} fading channels with the mean as one are modeled for radar and communication channels \cite{small}. The probability density function (PDF) can be expressed as ${f_{{{\left| {{h_i}} \right|}^2}}}\left( x \right) = \frac{{{m^m}}}{{\Gamma \left( m \right)}}{x^{m - 1}}\exp \left( { - mx} \right)$ with $m$ as the Nakagami-\emph{m} shape parameter. As shown in Fig. \ref{system}, we denote $i = \left\{ {\left( {r,d} \right),\left( {r,u} \right),r,c} \right\}$ to present different small-scale channel gains, i.e., ${\left| {{h_{r,d}}} \right|^2}$ and ${\left| {{h_{r,u}}} \right|^2}$ for the downlink transmission and uplink echoes of the radar target, ${{{\left| {{h_c}} \right|}^2}}$ for the communication transmitter's uplink signals, and ${{{\left| {{h_r}} \right|}^2}}$ for the radar target's uplink signals.

\subsubsection{Large-Scale Fading}
The path loss models are split into two categories. For the NOMA cluster to be evaluated, we define the distance between the BS and the communication transmitter as $d_c$. Additionally, the distance between the BS and the radar target is denoted as $d_r$. For the communication channels, the path loss function follows the conventional path loss model, which can be expressed as
 \begin{align}
{\mathcal{P}_c}\left( {{d_c}} \right) = {C_c}{\left( {{d_c}} \right)^{ - {\alpha _c}}},
\end{align}
where $\alpha _c$ is the path loss exponent for the communication links, ${C_c}={\left( {\frac{c}{{4\pi {f_c}}}} \right)^2}$ with the reference distance $d_0 = 1$ m, the speed of light $c=3\times10^8$ m/s, and the carrier frequency $f_c$.

For the channels of radar echoes, the BS that transmits pulse repetition intervals is considered. Including the downlink transmission and uplink echoes, the path loss function of the radar echoes can be presented as
\begin{align}
{\mathcal{P}_r}\left( {{d_r}} \right) = {C_r}{\left( {{d_r}} \right)^{ - {\alpha _r}}},
\end{align}
where $\alpha _r$ is the path loss exponent for the radar echoes, e.g., $\alpha _r=4$ for free-space scenarios presented in \cite{small}. The parameter ${C_r} = \frac{{\sigma_{RCS} {\lambda ^2}}}{{{{\left( {4\pi } \right)}^3}}}$ is the reference-distance-based intercept with $\lambda$ as the wavelength of carrier and $\sigma_{RCS}= \frac{{4\pi {S_r}}}{{{S_t}}}$ as the target radar cross section, where $S_r$ is the power density that is intercepted by the target and $S_t$ is the scattered power density in the reference range $d_0=1$ m \cite{small}.

\subsection{NOMA Semi-ISaC case}

 For the NOMA Semi-ISaC case, the total bandwidth is split as two parts, i.e., the ISaC bandwidth and the communication-only bandwidth. With the aid of the SIC technique, we utilize the power domain to achieve the multiple access instead of the frequency domain. Under a two-use NOMA Semi-ISaC case, a communication transmitter is paired with a radar target to form one NOMA cluster that utilizes the same orthogonal resource block. With the aid of SIC processes, the received signals, including the information of both radar and communication systems, are handled and decoded in sequence based on various power levels. Compared to the conventional NOMA system, the NOMA Semi-ISaC system under the ISaC bandwidth needs one more SIC process than the conventional NOMA system to obtain the radar signals.

 As the deployment of two users influences the SIC orders, we consider the scenario where a near communication transmitter is paired with a far radar target. The analysis of the other scenario, where a far communication transmitter is paired with a near radar target, can be obtained with the same processes of the first scenario, which is emitted because of the limitation of space.

\subsection{SIC Orders}

The SIC processes of conventional NOMA systems and ISaC NOMA systems are different. For conventional NOMA systems, the BS only receives signals from two different power levels when the two-user cases are considered. The SIC orders are noted that near users obtain better channel conditions than far ones, thus the nearest user is decoded at the first stage of SIC processes\footnote{With fixed power allocation, the path loss have more stable and dominant effects than small-scale fading as we consider the averaged performance. Thus, we consider the near users as the strong users}. For ISaC NOMA systems under the two-user case, the BS receives three parts of signals, i.e., the communication signals from the radar target, the communication signals from the transmitter, and the reflected echoes by the radar target. We note that pulse repetition intervals are transformed for the radar echoes, thus they have no modulation and coding process. Hence, the communication signals from the two paired users have higher priority than the radar echoes. In conclusion, it is better to fix the SIC order of radar echoes at the last stage. For the two communication signals, the near user's signals are decoded firstly and the far user's signals are decoded in the middle stage.

As we fix the SIC order of radar echoes at last, a drawback is obtained that when the radar echoes experience strong channels with large power levels, the ISaC system may face low performance as the echoes are regarded as interference for communication signals. To mitigate unnecessary interference from radar signals, we exploit the predicted target range to generate a predicted radar return and subtract it from the integrated signals \cite{ref}. By this approach, the performance of the communication system is improved. By subtracting the predicted target range, the jointly received signal model $v(t)$ is expressed as
\begin{align}
  v\left( t \right) =& \underbrace {{h_{r,d}}{h_{r,u}}\sqrt {{P_{BS}}{\mathcal{P}_r}\left( {{d_r}} \right)} \left[ {x\left( {t - \tau } \right) - x\left( {t - {\tau _{pre}}} \right)} \right]}_{{e_r}} \notag\\
  & + \underbrace {{h_c}\sqrt {{P_c}{\mathcal{P}_c}\left( {{d_c}} \right)} z\left( t \right)}_{{s_c}} + \underbrace {{h_r}\sqrt {{P_r}{\mathcal{P}_c}\left( {{d_r}} \right)} y\left( t \right)}_{{s_r}} + n\left( t \right),
\end{align}
where $s_c$ is the received communication signals from the communication transmitter to the BS, $s_r$ is the received communication signals from the radar target to the BS, and $e_r$ is the radar echoes from the radar target to the BS. Additionally, The $P_c$ and $P_r$ are the uplink transmit power of the communication transmitter and the radar target, respectively. The $P_{BS}$ is the transmit power from the BS for the radar detection. The $n\left( t \right)$ represents the noise with the strength $\sigma ^2=k_B T_{temp} B$, where $k_B$ is the Boltzmann constant and $T_{temp}$ is the absolute temperature. The $\tau$ is the observation time delay of the radar targets and $\tau_{pre}$ is the predicted value of $\tau$.

\subsection{Signal Model}

We assume that the BS has prior observations to evaluate the predicted range of radar target positions. Additionally, the position fluctuation is able to be presented by time delay fluctuation in radar systems. Hence, the time delay fluctuation $\tau$ is denoted as a Gaussian distribution, with the variance $\sigma_{\tau}^2=\mathbb{E}\left[ {{{\left| {\tau  - {\tau _{pre}}} \right|}^2}} \right]$, where $\mathbb{E}\left[ {\cdot} \right]$ is the expectation. We present the total bandwidth as $B$, thus the bandwidth used for the ISaC case is presented as $(\beta_{semi}B)$ with $\beta_{semi} \in [0,1]$. In this case, the expectation of time delay fluctuation, denoted as $E_{TD}$, can be derived as
\begin{align}
E_{TD} = \mathbb{E}\left[ {{{\left| {x(t - \tau ) - x(t - {\tau _{pre}})} \right|}^2}} \right] \approx \gamma^2 {\beta_{semi}^2} B^2 \sigma_{\tau}^2,
\end{align}
where $\gamma^2 = (2\pi)^2/12$ for a flat spectral shape. This solution is able to be derived by replacing the time delay as a derivative and employing Parserval's theorem with the Fourier transform \cite{ref}.

\subsubsection{Communication Signals}

With receiving different power levels, the BS directly decodes the communication signals of the communication transmitter by considering the communication signals and the radar echoes of the radar target as interference. Hence, the signal-to-interference-and-noise ratio (SINR) of the communication transmitter is expressed as
\begin{align}\label{S-I-c}
\gamma _c^I = \frac{{{P_c}{\mathcal{P}_c}\left( {{d_c}} \right){{\left| {{h_c}} \right|}^2}}}{{{P_r}{\mathcal{P}_c}\left( {{d_r}} \right){{\left| {{h_r}} \right|}^2} + {P_{BS}}{\mathcal{P}_r}\left( {{d_r}} \right){{\left| {{h_{r,d}}} \right|}^2}{{\left| {{h_{r,u}}} \right|}^2}E_{TD}  + {\sigma ^2}}},
 \end{align}

After removing the signals of the communication transmitter by the SIC process, the SINR of the communication signals for the radar target is presented as
\begin{align}\label{S-I-r}
\gamma _r^I {=} \frac{{{P_r}{\mathcal{P}_c}\left( {{d_r}} \right){{\left| {{h_r}} \right|}^2}}}{{{P_{BS}}{\mathcal{P}_r}\left( {{d_r}} \right){{\left| {{h_{r,d}}} \right|}^2}{{\left| {{h_{r,u}}} \right|}^2}E_{TD}  + {\sigma ^2}}}.
 \end{align}

\subsubsection{Radar Echoes}
As we would like to ensure the priority of communication signals, the radar echoes are fixed at the last stage of SIC orders. Hence, with the aid of the SIC process to cancellate all communication signals, the SNR is presented as
\begin{align}
\gamma _r^{echo} = \frac{{{P_{BS}}{\mathcal{P}_r}\left( {{d_r}} \right){{\left| {{h_{r,d}}} \right|}^2}{{\left| {{h_{r,u}}} \right|}^2}E_{TD} }}{{{\sigma ^2}}}.
\end{align}

For radar echoes, the performance of radar estimation is better when the SNR of the radar echoes is enlarged. In the following derivations (in Section III and IV), we will consider a metric called ergodic REIR to estimate the performance of the radar system. In this metric, the SNR is directly utilized.

\section{Communication Performance Evaluation}

Based on two scenarios with different user deployment, the outage performance of communication signals from NOMA users is investigated. We first derive the closed-form expressions with or without considering the random deployment of users, respectively. Additionally, the asymptotic outage performance and the diversity gains of users are evaluated.

When the communication signals are processed, the pulse intervals reflected by the radar target are considered as interference. We thus evaluate the averaged interference strength before analyzing the outage performance of users.

\begin{lemma}\label{AveragedInterference}
\emph{We denote ${I_R} = {P_{BS}}{\mathcal{P}_r}\left( {{d_r}} \right){\left| {{h_{r,d}}} \right|^2}{\left| {{h_{r,u}}} \right|^2}E_{TD} $ to simplify the expression of interference (radar echoes). The expectation of interference is derived as }
\begin{align}
\mathbb{E}\left[ {{I_R}} \right]\left( {{d_r}} \right) = {P_{BS}}{\mathcal{P}_r}\left( {{d_r}} \right){\gamma ^2}{\beta_{semi}^2}{B^2}\sigma _\tau ^2,
\end{align}
\emph{where $\mathbb{E}\left[ {{X}} \right]\left( {{x}} \right)$ is the expectation of $X$ with the variable $x$.}
\begin{IEEEproof}
With the definition of expectation and the distribution of Nakagami-\emph{m} fading channels, the expression of interference is presented as two integrations as
\begin{align}
  &\mathbb{E}\left[ {{I_R}} \right]\left( {{d_r}} \right) = {P_{BS}}{\mathcal{P}_r}\left( {{d_r}} \right){\gamma ^2}{\beta_{semi}^2}{B^2}\sigma _\tau ^2{\left( {\frac{{{m^m}}}{{\Gamma \left( m \right)}}} \right)^2} \notag \\
 &\hspace*{1cm}\times \int_0^\infty  {{x^m}\exp \left( { - mx} \right)dx} \int_0^\infty  {{y^m}} \exp \left( { - my} \right)dy,
\end{align}
and with the Eq. [2.3.3.1] in \cite{table}, this lemma is proved.
\end{IEEEproof}
\end{lemma}

\subsection{Outage Performance for Communication Signals}

The BS decodes the communication transmitter's signals by considering the signals of the radar target as interference. To decode the signals of the radar target, the BS has the SIC technique to eliminate unexpected signals. Thus far, the expressions of outage probability for the NOMA users is presented based on the SINR expressions \eqref{S-I-c} and \eqref{S-I-r} as
\begin{align}
\label{OP-I-c}
\mathbb{P}_c^I &= \Pr \left\{ {\gamma _c^I < {\gamma _{th}}} \right\},\\
\label{OP-I-r}
\mathbb{P}_r^I &= 1 - \Pr \left\{ {\gamma _c^I > {\gamma _{SIC}},\gamma _r^I > {\gamma _{th}}} \right\},
\end{align}
where $\gamma _{SIC}$ is the threshold of the SIC process and $\gamma _{th}$ is the threshold of signal transmission.

\subsubsection{Communication Signals of the Communication Transmitter}
In terms of the outage probability the communication transmitter, we substitute \eqref{S-I-c} into \eqref{OP-I-c} with an averaged interference of radar pulses as \textbf{Lemma \ref{AveragedInterference}}. The closed-form expression of outage probability is then derived via \textbf{Theorem \ref{OP_I_c_f}}.

\begin{theorem}\label{OP_I_c_f}
\emph{For the communication signals of the NOMA-assisted Semi-ISaC system, the closed-form expression of outage probability for the communication transmitter is derived as }
\begin{align}
 & \mathbb{P}_c^I {=} 1 {-} \exp \left( { {-} \frac{{m{\gamma _{th}}}}{{{P_c}}}\left( {{a_1} {+} {a_2}} \right)} \right)\sum\limits_{p = 0}^{m - 1} {\sum\limits_{r = 0}^p {\frac{{{m^r}\gamma _{th}^pC_p^r{{\left( {{P_r}{a_3}} \right)}^{p - r}}}}{{\left( {m - 1} \right)!p!}}} }  \notag \\
 &\times \frac{{{{\left( {{a_1} {+} {a_2}} \right)}^r}}}{{P_c^p}}\Gamma \left( {m {+} p {-} r} \right){\left( {\frac{{{\gamma _{th}}{a_3}{P_r}}}{{{P_c}}} {+} 1} \right)^{{ -} \left( {m + p - r} \right)}},
\end{align}
\emph{where ${a_1} = \frac{{{P_{BS}}{G_r}{C_r}{{\left( {{d_r}} \right)}^{ - {\alpha _r}}}{\gamma ^2}{\beta_{semi}^2}{B^2}\sigma _\tau ^2}}{{{G_c}{C_c}{{\left( {{d_c}} \right)}^{ - {\alpha _c}}}}}$, ${a_2} = \frac{{{\sigma ^2}}}{{{G_c}{C_c}{{\left( {{d_c}} \right)}^{ - {\alpha _c}}}}}$, ${a_3} = \frac{{{{\left( {{d_r}} \right)}^{ - {\alpha _c}}}}}{{{{\left( {{d_c}} \right)}^{ - {\alpha _c}}}}}$, and ${C_n^m}=n!/(m!(n-m)!)$. Additionally, we note that $\Gamma(x)$ is the Gamma function.}
\begin{IEEEproof}
See Appendix~A.
\end{IEEEproof}
\end{theorem}

\subsubsection{Communication Signals of the Radar Target}
Considering the averaged interference of radar echoes, the outage probability expressions are obtained by substituting \eqref{S-I-c} and \eqref{S-I-r} into \eqref{OP-I-r}. We evaluate the outage performance of the communication signals of the radar target as \textbf{Theorem \ref{OP_I_r_f}}.

\begin{theorem}\label{OP_I_r_f}
\emph{For the communication signals of the radar target, the closed-form expression of the outage probability is derived as }
\begin{align}
  \mathbb{P}_r^I &= 1 - \sum\limits_{p = 0}^{m - 1} {\sum\limits_{r = 0}^p {C_p^r} \frac{{{{\left( {{a_1} + {a_2}} \right)}^{p - r}}{{\left( {{a_3}{P_r}} \right)}^r}}}{{\Gamma \left( m \right){m^r}p!}}} {\left( {\frac{{m{\gamma _{SIC}}}}{{{P_c}}}} \right)^p} \notag \\
  &\times\exp \left( { - \frac{{m{\gamma _{SIC}}\left( {{a_1} + {a_2}} \right)}}{{{P_c}}}} \right){\left( {\frac{{{\gamma _{SIC}}{a_3}{P_r}}}{{{P_c}}} + 1} \right)^{ - (r + m)}} \notag \\
  & \times{\Gamma \left( {r {+} m,\frac{{{\gamma _{th}}m\left( {{a_4} {+} {a_5}} \right)}}{{{P_r}}}\left( {\frac{{{\gamma _{SIC}}{a_3}{P_r}}}{{{P_c}}} + 1} \right)} \right)},
\end{align}
\emph{where ${a_4} = \frac{{{P_{BS}}{G_r}{C_r}{{\left( {{d_r}} \right)}^{ - {\alpha _r}}}{\gamma ^2}{\beta_{semi}^2}{B^2}\sigma _\tau ^2}}{{{G_c}{C_c}{{\left( {{d_r}} \right)}^{ - {\alpha _c}}}}}$, ${a_5} = \frac{{{\sigma ^2}}}{{{G_c}{C_c}{{\left( {{d_r}} \right)}^{ - {\alpha _c}}}}}$, and $\Gamma \left( {\cdot,\cdot} \right)$ is the upper incomplete Gamma function.}
\begin{IEEEproof}
See Appendix~B.
\end{IEEEproof}
\end{theorem}

\section{Radar Performance Evaluation}
Based on the former section, we have evaluated the communication signals of two NOMA users when considering the radar echoes as interference. In this section, we will focus on the performance of the radar echoes. We propose a new performance matric called ergodic \emph{Radar Estimation Information Rate (REIR)}, which is the ergodic process of the RIER to evaluate the averaged performance of the radar echoes. Considering the entropy of a random parameter and the rate distortion theory, the former technical research based on \cite{ref} has proposed the RIER, which is analogous to data information rate for instantaneous performance. This REIR is able to be further simplified by the Cram\'{e}r-Rao lower bound, which has a clear relationship between the REIR and the transmit SNR, which is presented as
\begin{align}
{R_{est}} \le \frac{\delta }{{2T}}{\log _2}\left( {1 + 2T{\beta_{semi}}B\gamma _r^{echo}} \right),
\end{align}
where $\gamma _r^{echo} = \frac{{{P_{BS}}{\mathcal{P}_r}\left( {{d_r}} \right){{\left| {{h_{r,d}}} \right|}^2}{{\left| {{h_{r,u}}} \right|}^2}{\gamma ^2}{\beta_{semi}^2}{B^2}\sigma _\tau ^2}}{{{\sigma ^2}}}$ is the signal to noise ratio for the radar echoes of the radar target and $\delta$ is the duty factor (duty cycle) of the radar target. Hence, after the SIC process, we obtain the SNR of the radar echoes and substitute it into the equation to derive our ergodic REIR in the following.

\subsection{Equivalent Radar Channels}
The radar transmitting channel is considered as two independent links, i.e., a downlink transmission and a reflecting link from the radar target. Thus, the equivalent small scale channel fading is able to be expressed as ${\left| {{h_{r,eq}}} \right|^2} = {\left| {{h_{r,d}}} \right|^2}{\left| {{h_{r,u}}} \right|^2}$. We first present the equivalent channel distribution of ${\left| {{h_{r,eq}}} \right|^2}$ and then derive the ergodic REIR.

\begin{lemma}\label{channel_lemma}
\emph{When we consider the two links are Nakagami-m fading channels, the PDF and CDF of the equivalent channel is derived in the following.}
\begin{align}
  {f_{{{\left| {{h_{r,eq}}} \right|}^2}}}\left( z \right) &= \frac{{2{m^{2m}}}}{{{{\left( {\Gamma \left( m \right)} \right)}^2}}}{z^{m - 1}}{K_0}\left( {2m\sqrt z } \right) ,\\
  {F_{{{\left| {{h_{r,eq}}} \right|}^2}}}\left( z \right) &= \frac{{G{_1^2}{_3^1}\left( {{m^2}x\left| {_{m,m,0}^1} \right.} \right)}}{{{{\left( {\Gamma \left( m \right)} \right)}^2}}} ,
\end{align}
\emph{Where ${K_0}\left(  \cdot  \right)$ is the modified Bessel function of the third kind and ${G{_p^m}{_q^n}\left( {\cdot \left| {_{\left( {{b_q}} \right)}^{\left( {{a_p}} \right)}} \right.} \right)}$ is the Meijer G function.}
\begin{IEEEproof}
With the aid of ${K_v}\left( x \right) {=} \frac{1}{2}G{_0^2}{_2^0}\left( {\frac{{{x^2}}}{4}\left| {{_{\frac{v}{2}}^ \cdot}{ {_{\frac{ - v}{2}}^ \cdot}} } \right.} \right)$, $\int_0^x {{z^{m - 1}}G{_0^2}{_2^0}\left( {{m^2}z\left| {{_0^ \cdot}{ _0^ \cdot} } \right.} \right)} dz {=} {x^m}G{_1^2}{_3^1}\left( {{m^2}y\left| {{_{0,0}^{1 - m}}{_{ - m}^ \cdot} } \right.} \right)$, $ {z^p}G{_p^m}{_q^n}\left( {z\left| {_{\left( {{b_q}} \right)}^{\left( {{a_p}} \right)}} \right.} \right) {=} {z^p}G{_p^m}{_q^n}\left( {z\left| {_{\left( {{b_q}} \right) + p}^{\left( {{a_p}} \right) + p}} \right.} \right)$, and Eq.[2.3.6.7] in \cite{table}, the above PDF and CDF are able to be derived.
\end{IEEEproof}
\end{lemma}

\subsection{Ergodic Radar Estimation Information Rate}

Based on the equivalent channel distribution, we derive the ergodic REIR of radar signals in \textbf{Theorem \ref{EREIR1}}. As the \textbf{Theorem \ref{EREIR1}} has no closed-form expression, we release the constraint as $m \in Z$ in \textbf{Corollary \ref{EREIR2}} and $m =1 $  \textbf{Corollary \ref{EREIR3}}.

\begin{theorem}\label{EREIR1}
\emph{Based on the channel models in \textbf{Lemma \ref{channel_lemma}}, the expression of the ergodic REIR for the radar target is derived as }
\begin{align}
{R_{est}} {\approx} \frac{\delta }{{2T\ln \left( 2 \right)}}\int_0^\infty  {\frac{1}{{z{ +} 1}}\left( {1 {-} \frac{{G{_1^2}{_3^1}\left( {\frac{{{m^2}d_t^{{\alpha _r}}}}{{{\Xi _{r,1}}}}z\left| {_{m,m,0}^1} \right.} \right)}}{{{{\left( {\Gamma \left( m \right)} \right)}^2}}}} \right)} dz.
\end{align}
\begin{IEEEproof}
With the aid of \textbf{Lemma 2}, the theorem is proved.
\end{IEEEproof}
\end{theorem}

\begin{corollary}\label{EREIR2}
\emph{Considering the special case with $m \in Z$, }
\begin{align}
{R_{est}} {\approx} \sum\limits_{k {=} 0}^{m {-} 1} {\frac{\delta \int_0^\infty  {\exp \left( {\frac{{md_t^{{\alpha _r}}}}{{{\Xi _{r,1}}x}}} \right){E_{k + 1}}\left( {\frac{{md_t^{{\alpha _r}}}}{{{\Xi _{r,1}}x}}} \right){f_{{{\left| {{h_{r,u}}} \right|}^2}}}\left( x \right)dx}}{{2T\ln \left( 2 \right)}}}  ,
\end{align}
\emph{where ${E_n}\left( \cdot \right)$ is the generalized exponential integral with the parameter $n$.}
\begin{IEEEproof}
The corollary is proved by utilizing the relationship between ${E_n}\left( \cdot \right)$ and the Meijer G function.
\end{IEEEproof}
\end{corollary}

\begin{corollary}\label{EREIR3}
\emph{Under the special case with $m = 1$, which means the channels of radar targets are Rayleigh fading channels, the closed-form expression of the ergodic REIR is derived as }
\begin{align}
{R_{est}} \approx \frac{\delta }{{2T\ln \left( 2 \right)}}G{_1^3}{_3^1}\left( {d_t^{{\alpha _r}}\Xi _{r,1}^{ - 1}\left| {_{0,0,1}^0} \right.} \right).
\end{align}
\begin{IEEEproof}
This corollary is proved by exploiting Eq.[2.3.4.4] in \cite{table} and the Meijer G function.
\end{IEEEproof}
\end{corollary}

To derive the closed-from expressions of the REIR, we need some approximated expressions for special functions, i.e., $\gamma \left( {m,t} \right) = \left( {m - 1} \right)! - \exp \left( { - t} \right)\sum\limits_{k = 0}^{m - 1} {\frac{{\left( {m - 1} \right)!}}{{k!}}{t^k}} $, ${E_n}\left( z \right) \approx \frac{{{{\left( { - z} \right)}^{n - 1}}}}{{\left( {n - 1} \right)!}}\left( {\psi \left( n \right) - \ln \left( z \right)} \right) - \sum\limits_{k = 0\& k \ne n - 1}^\infty  {\frac{{{{\left( { - z} \right)}^k}}}{{k!\left( {1 - n + k} \right)}}} $ for $n>1$, and ${E_1}\left( z \right) \approx  - {C_\gamma } - \ln \left( z \right) + z$, where ${C_\gamma }$ is the Euler constant and ${\psi \left( n \right)}$ is the Psi function. Because of the limitation of space, the derivations are emitted.

\section{Numerical Results}

Numerical analysis is presented in this section. Without specific notions, the value of coefficients are set as: the distance of the near user as 800 meter and that of the far user as 1300 meter, the bandwidth $B=10$ MHz, the noise power $\sigma^2= k_b B T_{temp}$ with $T_{temp}=724$ K, the threshold for communication $\gamma_{th}=1$, the threshold for SIC $\gamma_{SIC}=0.4$, the carrier frequency $f_c = 10^9$ Hz, the speed of light $c = 3*10^8$ m/s, the target radar cross section $\sigma_{RCS}=0.1$, the pulse duration $T=1$ $\mu$s, the path loss exponents $\alpha_r=4.5$ and $\alpha_c=2.5$, the radar duty factor $\delta = 0.01$, and the Nagakami coefficient $m=3$.

\begin{figure}[!htb]
\centering
\includegraphics[width= 3in]{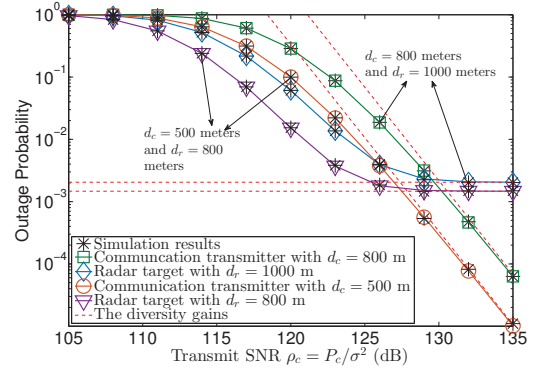}
\caption{Outage probability versus the transmit SNR of the communication transmitter $\rho_c$. }
\vspace{-0.3cm}
\label{figure1}
\end{figure}

We validate the analytical of the outage probability for NOMA users by Fig. \ref{figure1}. Based on the simulation results, the analytical results match the simulation results well. One conclusion is that when we enlarge the near user's transmit SNR, the performance of two users will be enhanced while the far user has the lower limit. The reason is explained in the following. For the near user, enhancing its transmit SNR directly increases the strength of signals, resulting in reduced outage probability. For the far user, enhancing the near user's transmit SNR is only able to increase the success rate of SIC. When the SIC process is perfect, the lower limit comes.

\begin{figure}[!htb]
\centering
\includegraphics[width= 3in]{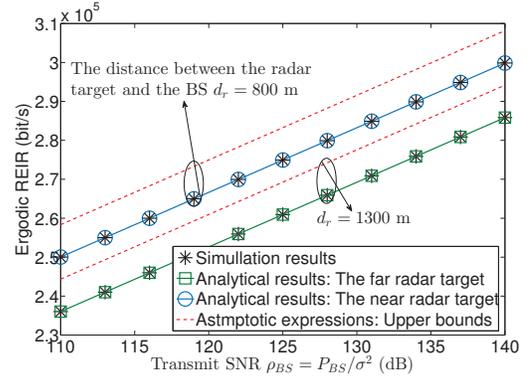}
\caption{The ergodic rate versus the transmit SNR of the BS $\rho_{BS}$ with various distance $d_r = [800,1300]$ meters.}
\vspace{-0.3cm}
\label{figure2}
\end{figure}
\begin{figure}[!htb]
\centering
\includegraphics[width= 3in]{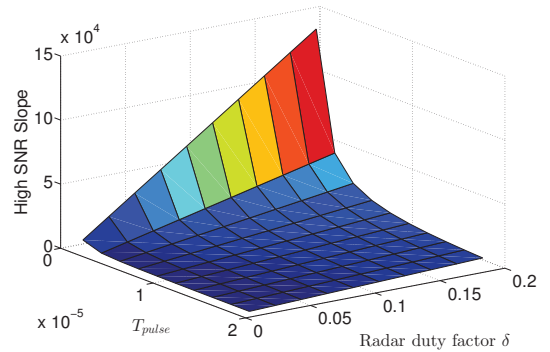}
\caption{The ergodic rate varies based on different value of the duty factor $\delta$ and the pulse duration $T_{pulse}$.}
\vspace{-0.3cm}
\label{figure3}
\end{figure}

In Fig. \ref{figure2}, the expressions of the ergodic REIR are validated. The analytical results fit the simulation results well. We conclude that the ergodic REIR is enlarged if the pulses are dense. This is because the total signals transformed by the BS are increased with dense pulses. In Fig. \ref{figure3}, it represents the diversity gains of ergodic REIR, defined as $S = \mathop {\lim }\limits_{{P_{BS}} \to \infty } \frac{{R_{est}\left( {{P_{BS}}} \right)}}{{\log \left( {{P_{BS}}} \right)}}$. Firstly, it is indicated that the user with a higher radar duty factor has a better ergodic rate since increasing the radar duty factor means to increase the radar pulse duration in the same period. Additionally, one is concluded that with the same radar duty factor, the user with lower pulse duration has a better ergodic rate. This is because when the radar duty factor is fixed, reducing the pulse duration is able to reduce the waiting time between two pulses, thus the density of pulses is enlarged.

\section{Conclusion}

We have proposed the Semi-ISAC network, where the total bandwidth is split into two portions, i.e., the communication-only bandwidth and the ISaC bandwidth. When decoding the communication signals, we consider the radar echoes as interference but only the ISaC bandwidth has such interference. We have defined a new performance metric, i.e., ergodic REIR, to evaluate the averaged radar estimation rate. We have first derived the expressions of outage probability for communication signals. Additionally, for radar echoes, we have derived the expressions of ergodic REIR. The analysis of this paper has verified that: 1) dense pulses can enhance the performance of radar target, while this will enlarge the interference for the communication signals; 2) The transmit power of the BS can be adjusted to reduce the interference of radar signals; and 3) To obtain optimized performance, we can balance the radar and communication signals by adjusting the transmit power of the BS and the frequency of pulses together.

\section*{Appendix~A: Proof of Theorem~\ref{OP_I_c_f}} \label{Appendix:A}
\renewcommand{\theequation}{A.\arabic{equation}}
\setcounter{equation}{0}

To address the closed-form expressions of outage probability for the communication transmitter, the probability expression should be manipulated as $
\mathbb{P}_c^I = \Pr \left\{ {\gamma _c^I < {\gamma _{th}}} \right\}$. Substituting the expectation of interference in \textbf{Lemma \ref{AveragedInterference}} and rewriting the probability equation as integrations, we present the outage probability expression by exploiting the PDF and CDF of Nakagami-\emph{m} fading channels as
\begin{align}\label{A1}
\mathbb{P}_c^I = \int_0^\infty  {\frac{1}{{\Gamma \left( m \right)}}\gamma \left( {m,m\Lambda } \right)} {f_{{{\left| {{h_r}} \right|}^2}}}\left( x \right)dx,
\end{align}
where $\Lambda  = \frac{{{\gamma _{th}}{P_r}x{{\left( {{d_r}} \right)}^{ - {\alpha _c}}}}}{{{P_c}{{\left( {{d_c}} \right)}^{ - {\alpha _c}}}}} + \frac{{{\gamma _{th}}\mathbb{E}\left[ {{I_R}} \right]\left( {{d_r}} \right) + {\gamma _{th}}{\sigma ^2}}}{{{P_c}{G_c}{C_c}{{\left( {{d_c}} \right)}^{ - {\alpha _c}}}}}$.

As the CDF of Nakagami-\emph{m} fading channel (in power domain) is a lower incomplete Gamma function, one accurate series of the incomplete Gamma function is able to be exploited to reduce the complexity of derivation, which is expressed as
\begin{align}
\gamma \left( {a,b} \right) &= \Gamma \left( a \right) - \Gamma \left( {a,b} \right) \notag\\
&= \Gamma \left( a \right) - \sum\limits_{p = 0}^{a - 1} {\frac{{\left( {a - 1} \right)!}}{{p!}}} \exp \left( { - b} \right){b^p},
\end{align}
where $\Gamma \left( {a,b} \right)$ is the upper incomplete Gamma function.

And by substituting this equation into \eqref{A1} and with the aid of Binomial theorem, the former expression \eqref{A1} is further derived as
\begin{align}
  &\mathbb{P}_c^I {=} 1 {-} \exp \left( { {-} \frac{{m{\gamma _{th}}\left( {{a_1} + {a_2}} \right)}}{{{P_c}}}} \right)\sum\limits_{p = 0}^{m - 1} {\frac{{{{\left( {{a_1} + {a_2}} \right)}^r}{{\left( {{P_r}{a_3}} \right)}^{p - r}}}}{{p!}}}  \notag \\
  &\times\sum\limits_{r = 0}^p {\frac{{C_p^r{{\left( {m{\gamma _{th}}} \right)}^p}\int_0^\infty  {\exp \left( { - \frac{{m{\gamma _{th}}{a_3}{P_r}}}{{{P_c}}}x} \right)} {x^{p - r}}{f_{{{\left| {{h_r}} \right|}^2}}}\left( x \right)dx}}{{P_c^p}}},
\end{align}
and we use Eq. [2.3.3.1] in \cite{table} to obtain the final expression.

\vspace{0.2cm}
\section*{Appendix~B: Proof of Theorem~\ref{OP_I_r_f}} \label{Appendix:B}
\renewcommand{\theequation}{B.\arabic{equation}}
\setcounter{equation}{0}

Under the case with the fixed near communication transmitter and the fixed far radar target, the outage probability for the radar target is expressed as
 \begin{align}
  \mathbb{P}_r^I =& 1 - \Pr \left\{ {{{\left| {{h_c}} \right|}^2} > {\gamma _{SIC}}\frac{{{a_3}{P_r}{{\left| {{h_r}} \right|}^2} + {a_1} + {a_2}}}{{{P_c}}},} \right. \notag \\
  &\left. {{{\left| {{h_r}} \right|}^2} > \frac{{{\gamma _{th}}\left( {{a_4} + {a_5}} \right)}}{{{P_r}}}} \right\} .
 \end{align}

We substitute the CDF of Nakagami-\emph{m} distribution to derive the probability expression as
\begin{align}
 & \mathbb{P}_r^I = 1 - \int_{\frac{{{\gamma _{th}}\left( {{a_4} + {a_5}} \right)}}{{{P_r}}}}^\infty  {{f_{{{\left| {{h_r}} \right|}^2}}}\left( x \right)}\times  \notag \\
 & \left( {1 - \frac{1}{{\Gamma \left( m \right)}}\gamma \left( {m,\frac{{m{\gamma _{SIC}}}}{{{P_c}}}\left( {{a_3}{P_r}{{\left| {{h_r}} \right|}^2} + {a_1} + {a_2}} \right)} \right)} \right)dx.
\end{align}

Using the series of the lower incomplete Gamma function, the outage probability expression is derived as
\begin{align}
  \mathbb{P}_r^I = &1 - \sum\limits_{p = 0}^{m - 1} {\frac{1}{{p!}}{{\left( {\frac{{m{\gamma _{SIC}}}}{{{P_c}}}} \right)}^p}\exp \left( { - \frac{{m{\gamma _{SIC}}\left( {{a_1} + {a_2}} \right)}}{{{P_c}}}} \right)}  \notag \\
   &\times \underbrace {\int_{\frac{{{\gamma _{th}}\left( {{a_4} + {a_5}} \right)}}{{{P_r}}}}^\infty  {\exp \left( { - \frac{{m{\gamma _{SIC}}{a_3}{P_r}x}}{{{P_c}}}} \right)} {x^r}{f_{{{\left| {{h_r}} \right|}^2}}}\left( x \right)dx}_{{I_3}}\notag\\
  &\times \sum\limits_{r = 0}^p {C_p^r} {\left( {{a_1} + {a_2}} \right)^{p - r}}{\left( {{a_3}{P_r}} \right)^r}.
\end{align}

And then, we derive $I_3$ based on Eq. [2.3.6.6] in \cite{table} as
  \begin{align}
  {I_3} =& \Gamma \left( {r + m,\frac{{{\gamma _{th}}m\left( {{a_4} + {a_5}} \right)}}{{{P_r}}}\left( {\frac{{{\gamma _{SIC}}{a_3}{P_r}}}{{{P_c}}} + 1} \right)} \right) \notag\\
  & \times \frac{1}{{\Gamma \left( m \right){m^r}}}{\left( {\frac{{{\gamma _{SIC}}{a_3}{P_r}}}{{{P_c}}} + 1} \right)^{ - (r + m)}}.
  \end{align}

  Finally, substitute $I_3$ into the outage probability expression, we can obtain the closed-form expression.

\vspace{0.2cm}

\bibliographystyle{IEEEtran}
\bibliography{mybib}

\end{document}